# Correlation between scale-invariant normal state resistivity and superconductivity in an electron-doped cuprate

Tarapada Sarkar[1], P. R. Mandal[1], N. R. Poniatowski[1], M. K. Chan[2] and Richard L. Greene[1]

[1]*Center for Nanophysics & Advanced Materials and Department of Physics, University of Maryland, College Park, Maryland 20742, USA.*
[2]*The National High Magnetic Field Laboratory, Los Alamos National Laboratory, Los Alamos, New Mexico 87545, USA.*

**An understanding of the normal state in the high-temperature superconducting cuprates is crucial to the ultimate understanding of the long-standing problem of the origin of the superconductivity itself. This so-called "strange metal" state is thought to be associated with a quantum critical point (QCP) hidden beneath the superconductivity[1,2]. In electron-doped cuprates—in contrast to hole-doped cuprates—it is possible to access the normal state at very low temperatures and low magnetic fields to study this putative QCP and to probe the T → 0 K state of these materials[3,4]. We report measurements of the low temperature normal state magnetoresistance (MR) of the n-type cuprate system $La_{2-x}Ce_xCuO_4$ (LCCO) and find that it is characterized by a linear-in-field behavior, which follows a scaling relation with applied field and temperature, for doping (x) above the putative QCP (x= 0.14)[5]. This unconventional behavior suggests that magnetic fields probe the same physics that gives rise to the anomalous low-temperature linear-in-T resistivity[4]. The magnitude of the linear MR decreases as $T_c$ decreases and goes to zero at the end of the superconducting dome (x ~0.175) above which a conventional quadratic MR is found. These results show that there is a strong correlation between the quantum critical excitations of the strange metal state and the high-$T_c$ superconductivity.**

Quantum criticality has been a recurrent theme for attempting to understand the physics of the cuprates and other strongly correlated materials[1,2]. But, despite extensive theoretical and experimental effort over the past 30 years, the relation between quantum criticality and the anomalous properties of the normal state and the origin of the superconductivity is unresolved. There has been much experimental evidence for a QCP as a function of doping in both electron-doped[3,6] and hole-doped[1,7-9] cuprates. However, the nature of the phase for the doping below the QCP is undetermined. For the n-type, the QCP is most likely associated with long range or short range antiferromagnetic order (AFM) (with the carrier $\ell_{mfp}$ < magnetic correlation length), while for the p-type the QCP marks the end of a pseudogap phase of unknown origin. The QCP in both cuprate types is associated with a Fermi surface reconstruction (FSR), where a large hole-like FS is found above the FSR doping[3,6,10-13] and a reconstructed FS of electron and hole pockets is found at lower doping. Remarkably, the normal state transport properties near the QCP are quite different for n- and p-type cuprates. Above Tc (for H =0) hole-doped cuprates exhibit the mysterious linear-in-T resistivity, which extends to high temperatures beyond the Mott-Ioffe-Regel (MIR)[14] limit (so-called "bad metal" behavior). Above $T_c$ ($H = 0$), the electron-doped cuprates exhibit an equally mysterious $\rho \sim T^2$ behavior extending to high temperature (800-1000 K)[15]. A recent paper[16] has discussed this anomalous $T \gg T_C$ resistivity in n-type and concluded that a non-Fermi liquid (FL) scattering related to strong interaction-induced hydrodynamics (driven by the underlying quantum criticality) is likely the cause. A concurrent study of the $T \gg T_C$ thermal diffusivity of n-type cuprates[17] has also suggested that this high-temperature (above ~ 250K) transport is of hydrodynamic origin, i.e., non-quasiparticle transport of a fluid of electrons and phonons controlled by "Planckian" dissipation[18,19].

The role of a magnetic field on the QCP and the normal state properties is also undetermined in spite of much theoretical and experimental effort. The electron-doped cuprates have a much lower critical field ($H_{c2}$ < 10T)[3,20] than hole-doped cuprates and this allows access to the very low-temperature (T→0K) normal state properties. For example, Hall Effect studies at 400 mK have suggested that the FSR occurs at a doping just above the doping for maximum Tc (optimal doping); 0.17 in $Pr_{2-x}Ce_xCuO_4$ (PCCO)[6] and 0.14 in LCCO[5], a conclusion confirmed by ARPES[3,21] and quantum oscillation experiments for PCCO[22] and $Nd_{2-x}Ce_xCuO_4$ (NCCO)[12]. Also, for fields just above $H_{c2}$ it was shown that a linear-in-T resistivity extends from 10K down to 30 mK in LCCO for the doping range 0.14 to 0.17 and that the strength of this linear term correlates with the magnitude of $T_c$[4]. The linear-in-T resistivity---in contrast to the expectation of Fermi liquid quadratic temperature dependence at such low temperatures---suggests a strong interaction between the metallic electrons and the critical fluctuations associated with the QCP.

What has not been studied (or understood) is how the magnetic field may impact these fluctuations and hence the low temperature metallic state. In this work we remedy this deficiency by measurements of the MR in the normal state of LCCO in the same temperature and doping range (0.14 < x < 0.17) where the linear-in-T resistivity is found. We find a linear-in-H behavior, a distinctly different response than for quasiparticles in conventional metals, where one expects a MR ~$H^2$ for fields where $\omega_c\tau$ <<1. *The magnitude of the linear-in-H resistivity mirrors the magnitude and doping evolution of the linear-in-T resistivity, with both going to zero at the end of the superconducting dome.* Moreover, the temperature dependent magnetoresistance follows a scaling relation with applied field and temperature for doping (x) above the QCP up to the end of the superconducting dome. This shows that there are excitations, common to both field and temperature, which are correlated with the superconductivity (and probably the cause).

In Fig. 1 (a), (b) and (c) we show the *ab* plane magnetoresistance (MR) at temperatures between 400 mK and 15 K for c-axis field up to 14 T, for $La_{2-x}Ce_xCuO_4$ (LCCO) thin films with doping *x*= 0.15, 0.16 and 0.17. At low temperatures the normal state MR is linear-in-field for all doping. A linear-in-H to quadratic-in-H crossover occurs at higher temperatures at low field (<15 T). In the Fig. 4 we show *ab* plane magnetoresistance (MR) at temperatures between 360 mK and 60 K for c-axis dc field up to 31 T for doping x=0.15 and 0.16. In the SI, Fig S5 shows the *ab*-plane MR for a second sample with 0.15 doping measured in pulsed field up to 65 T. The MR measured at the lowest temperature fits well with $\Delta R \alpha \mu_0 H$. These high field, unsaturated, linear-in-*H* data strongly suggests that the low field (up to 14 T, Fig1a) MR is not a SC fluctuation effect. The linear-in-*H* data may continue to lower field, but the onset of superconductivity rules out a study of the normal state transport properties for $H < H_{c2}$. However, the measured MR just above the transition temperature strongly suggests linear-in-H resistivity continues to a much lower field (see Fig.4 and Figure S1).

In Fig. 2 we show the linear-in-temperature scattering coefficient $A(x)$ and the linear-in-field scattering coefficient $C(x)$ obtained from fits to the low temperature linear regions with $\rho(T) = \rho_0 + A(x)T$ and $\rho(H) = \rho(0) + C(x)(\mu_0 H)$. Both $A(x)$ and $C(x)$ decrease with $T_c$ as x increases and both go to zero at the doping where the superconductivity ends. In the non-superconducting overdoped regime the resistivity varies as $T^2$ (Fig.S9) and the MR goes as $H^2$ for T> 5K. A schematic temperature vs doping phase diagram in Fig.3b inset summarizes the temperature and field dependent magnetotransport data of LCCO.

These results are incompatible with that expected in conventional metals[23] where the MR from quasiparticles is controlled by the cyclotron frequency ($\omega_c = e\mu_0 H/m^*$) and the relaxation time, τ, i.e., $(\delta\rho/\rho(0))\sim(\omega_c\tau)^2 \propto H^2$ in the limit where $\omega_c\tau < 1$. At 14 T and 400 mK we estimate $\omega_c\tau < 0.15 \pm 0.05$ for our LCCO films. We do observe a MR proportional to $H^2$ at higher temperatures at low field (Figs 1, 4, S1, S2) where we also find[16] ρ proportional to $T^2$. This crossover in transport behavior as a function of T is shown schematically in Fig. S8. However, the high field (>20 T), high temperature, magnetoresistance remains linear with field as shown in Fig. 4. Our results suggest scale-invariant transport (i.e., lack of an intrinsic energy scale), which is often associated with quantum criticality. To check for this, we try the scaling analysis proposed in ref-24. In Fig. 3(a) we plot $\Delta\rho = (\bar{\rho} - \bar{\rho}(0))/T$ vs $\mu_0 H/T$ (where $\bar{\rho} = \frac{\rho(H, T)}{\rho(0, 200)}$ is normalized with $\rho(0,200)$ to avoid any geometrical error) in the normal state of the x=0.15 film. The data show a scaling with $\Delta\rho/T = (\alpha + \beta(\mu_0 H/T)^m$ (m=1.09 ± 0.01) where α and β are the fitting parameters. Taking m=1 at low temperatures (below <30 K), we can write $\Delta\rho = \alpha T + \beta\mu_0 H$. Converting to energy units, we write $\Delta\rho \propto (A(x)k_B T + C(x)\mu_B\mu_0 H) \equiv \varepsilon(T,H)$ with $\varepsilon(T, H)$, the sum of thermal energy and magnetic field energy. Figure 3(b) shows all the magneto transport data as a function of $\varepsilon(T, H)$, where $k_B$ is Boltzmann constant, $\mu_B$ is the Bohr magneton, $A (A/k_B = 2.3$ μΩ-cm/meV see Fig. S7) is the rate of change of resistivity as a function of temperature from Fig. 2b and $C$ ($C/\mu_B = 0.5$ μΩ − cm/meV at 0.4 K See Fig. S7) is the rate of change of resistivity as a function of magnetic field from Fig. 2a. This scaling is seen in x=0.16 and 0.17 as well (see Fig. S4). This means that the resistivity is linear with energy scale $\varepsilon(T,H)$ at low temperatures for doping between the putative QCP (x=0.14) and the end of the superconducting dome. This strongly suggests that the quantum critical region is scale invariant and that the H-linear behavior has the same origin as the T-linear behavior. The region of scale invariance is schematically shown in the Fig. 3(b) inset.

In LCCO we know that there is a Fermi surface reconstruction (FSR) at x = 0.14[5] presumably caused by the end of AFM order. If the fluctuations associated with a QCP are responsible for the anomalous low temperature ρ ~ T and Δρ ~ H in LCCO then our data suggests that there is a quantum critical region (not just a point) from x -= 0.14 to the end of the SC dome at $x_c$ ~0.175. As shown in Fig. 2, the scattering coefficient *A(x)* of temperature and scattering coefficient *C(x)* of magnetic field, obtained from fits to the linear temperature and magnetic field resistivity, decrease with $T_c$ as *x* is increased and approach zero at the end of the SC dome. This unique trend of the scattering coefficients strongly suggests that $T_c$ and the anomalous scattering are linked to each other. There are many proposed origins of a linear-in-H magnetoresistance when $\omega_c\tau < 1$[25-27]. The low temperature linear magnetoresistance is independent of temperature as shown in Figs. 1, 4 and supported by the scaling (Fig.3).

At higher temperatures, the magnitude of the $H^2$ MR decreases with temperature (see Fig.4 and S2). In contrast the high field magnetoresistance increases with temperature (see Fig.4). This unusual behavior is consistent with a breakdown of weak-field magnetotransport ($\delta\rho/\rho(0) \propto H^2$) at low temperatures near a QCP[26]. This model predicts linear-in-H behavior at the QCP, which is consistent with our experimental data. This model does not explain the high temperature high field linear magnetoresistance. However, this high field and high temperature behavior is consistent with a recent proposal for magnetoresistance in a disordered Marginal Fermi Liquid system[27].

We note that a linear-in-field MR has just been reported in the hole-doped cuprate $La_{2-x}Sr_xCuO_4$ (LSCO)[28], but for a rather different doping, magnetic field and temperature range than for our results reported here for n-type LCCO. The LSCO data are at higher fields and temperatures (where $\omega_c\tau \approx 1$ at 20 T [30]) and for doping below the putative QCP, in the region where the FS is reconstructed, and the normal state resistivity has an upturn at low temperatures[29]. Other magneto resistivity measurements on LSCO for doping above the QCP have found both T-linear and $T^2$ resistivity and an $H^2$ MR[31]. It is also important to mention that linear in H is not unusual at high field in disordered system[27,30]. However it is quite unusual to see temperature independent magnetoresistance at low temperatures and weak field. Thus, the question of whether the superconductivity in n- and p-type cuprates comes from similar normal states will have to await lower temperature normal state measurements for the p-type.

The linear-in-T resistivity in both p- and n-type cuprates has recently been attributed to Planckian dissipation[9], i.e., a maximum inelastic relaxation rate, $\frac{\hbar}{\tau}$, given by $k_BT$ [18,19]. This idea appears to be inconsistent with the resistivity behavior of LCCO and other n-type cuprates because the scattering rate goes well beyond the Planckian $k_BT$ limit above ~40 K[3,16]. The linear-in-T resistivity behavior is only found below ~40K in the n-type cuprates and increases roughly as $T^2$ above that temperature up to 400-800K[16]. In ref 18, Hartnoll states that Planckian (non-quasiparticle or hydrodynamic or incoherent) transport is expected to be manifested only at high temperature (where roughly the electron $l_{mfp}$(mean free path) ~lattice constant). In fact, evidence for this behavior has been found above ~250K in NCCO and $Sm_{2-x}Ce_xCuO_4$ (SCCO) crystals via thermal diffusivity measurements [17] and in LCCO from resistivity measurements [16]. To the authors knowledge there is no agreed upon prediction for the field dependence of the resistivity in the Planckian dissipation limit.

In summary, we have found an unconventional linear-in-field magnetoresistance at low temperature and low field in the electron doped cuprate LCCO. This behavior is found over an extended doping regime above the purported quantum critical point (i.e., the Fermi surface reconstruction doping). The magnitude of the linear-in-H resistivity mirrors the magnitude and doping evolution of the linear-in-T resistivity, with both going to zero at the end of the superconducting dome. An H/T scaling suggests that there is an energy scale, common to both field and temperature, which is linked to the superconductivity. *Our results represent an important and novel aspect of the ground state of the electron-doped cuprates.* Moreover, these new results suggest an anomalous quantum criticality in LCCO where the so-called "strange-metal" state of the cuprates can extend to very low temperatures and fields and over a wide range of doping.

**Acknowledgements:** This work is supported by the NSF under Grant No.DMR-1708334, the Maryland "Center for Nanophysics and Advanced Materials (CNAM). The National High Magnetic Field Laboratory is supported by the National Science Foundation Cooperative Agreement No. DMR-1157490, DMR-1644779, the state of Florida, and the U.S. Department of Energy. We thank Johnpierre Paglione, Nick Butch and Sankar Das Sarma for very helpful discussions and comments on the manuscript. We thank William Coniglio for help with 31 T measurement. We also thank Joshua Higgins for his advice and help with some of the measurements.

**Author contributions**: R.L.G conceived and directed the project. T.S performed the analysis. T.S, P.R.M and N. R. P prepared the samples and performed the measurement. T.S and M.K.C performed the high magnetic fields measurement. R.L.G and T.S wrote the manuscript and discussed with all other authors.

**Additional information:**
Supplementary information is available in the online version of the paper. Reprints and permissions information is available online at www.nature.com/reprints.
Correspondence and requests for materials should be addressed to R. L. G.

**Competing financial interests:**
The authors declare no competing financial interests.

**Figure Captions:**

**Figure 1: Magnetoresistance vs doping:** Magneto-resistivity for $La_{2-x}Ce_xCuO_4$ (LCCO) thin films with *x*= 0.15, 0.16 and 0.17. (a), (b) and (c) *ab*-plane transverse resistivity versus magnetic field ($H//c$ axis) as a function of temperature (color solid line) for all x.

**Figure 2: Doping dependent scattering rate:** (a) *ab* plane resistivity vs magnetic field ($H//c$ axis) for $La_{2-x}Ce_xCuO_4$ (LCCO) thin films with x= 0.15,0.16 and 0.17 at 400 mK fitted with $\rho(H) = \rho(0) + C(x)(\mu_0 H)$ (solid orange line) (b) ab plane resistivity vs Temperature (T) in the field driven normal state for x=0.15(8T), x=0.16(7T), x=0.17(6T) fitted with $\rho(T) = \rho(0) + A(x)T$ (solid orange line); (c) magnitude of  T resistivity (A) (red), magnitude of  $\mu_0 H$ (c) (blue) taken from figure (a) and (b) and normalized $T_c$ with respect to optimal $T_c$  (black) *vs* doping with respective statistical error of three sample for each doping (red, blue and black arrow).

**Figure 3: Scaling between field and temperature for doping x=0.15:** (a) $(\bar{\rho} - \bar{\rho}(0))/$ T (where $\bar{\rho} = \rho(T)/\rho(200K)$, and $\bar{\rho}(0) = \frac{\rho(0, 0.4)}{\rho(0,200)}$ taken from figure 2a) vs $\frac{\mu_0 H}{T}$. This plot has been deduced by varying temperature at fixed field and by varying field at fixed temperature (color solid lines) for doping x=0.15. This plot is fitted with $\Delta\rho = \alpha + \beta(\mu_0 H/T)^{\gamma}$ ($\gamma$=1.09) (blue dashed line); (b) $\bar{\rho}(T, H) - \bar{\rho}(0, 0)$ vs $T + \frac{C(x)\mu_B}{A(x)k_B}(\mu_0 H) \equiv \varepsilon(T, H)/A(x)k_B$ for all magnetoresistance data (black). The $\bar{\rho}(0, 0)$ is taken from extrapolating the zero field resistivity data down to T=0. The red is the resistivity in 8 T after subtracting the $\bar{\rho}(0, 8T)$. Inset: Phase diagram in the normal state for overdoped LCCO at low temperatures. The x-T plane (H=0) is the region where linear-in-T resistivity is seen. The x-H plane (T=0) is where linear magnetoresistance is seen at low temperatures. The $\varepsilon$ (T, H) is the energy scale below which linear resistivity is found (blue dashed line).

**Figure 4: High field Magnetoresistance:**

Transverse *ab* plane magnetoresistance MR for doping x= 0.15 (left panel-0.36 K (—), 1.6 K(—), 15 K(—), 25 K(—), 42 K(—), 60 K(—)) and 0.16 (right panel- 0.36 K(—), 1 K(—), 18 K(—), 25 K(—), 42 K(—), 60 K(—)) measured up to dc field of 31 T.

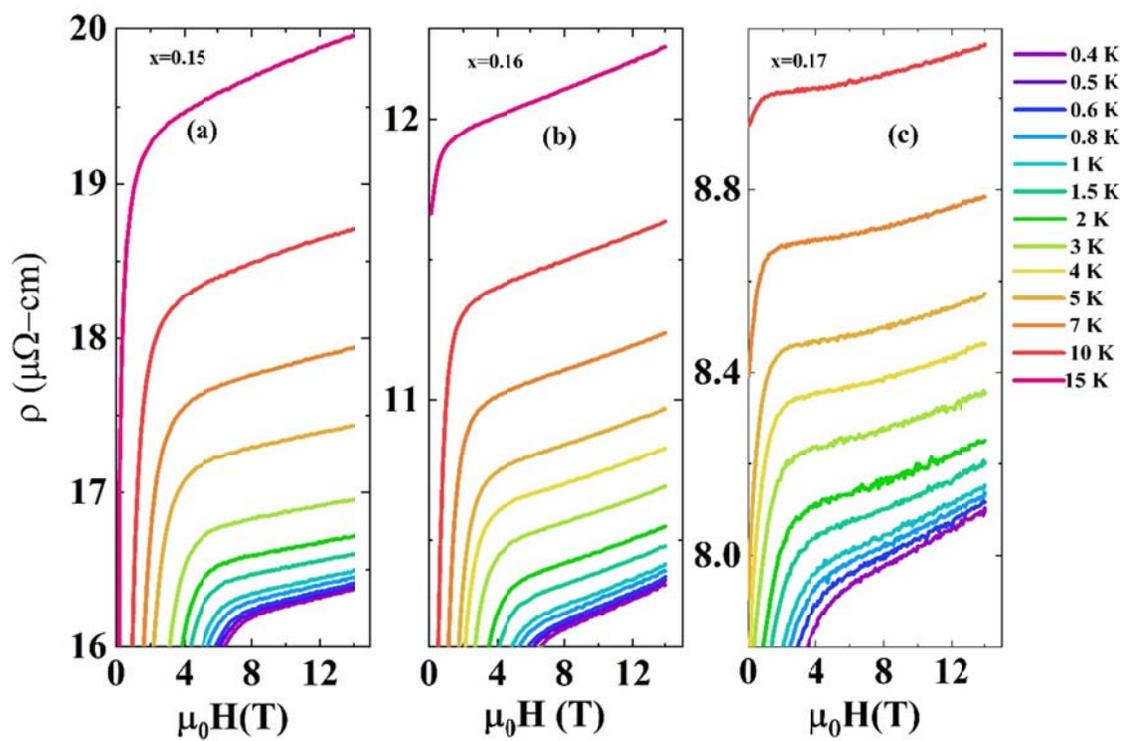

**Figure-1**

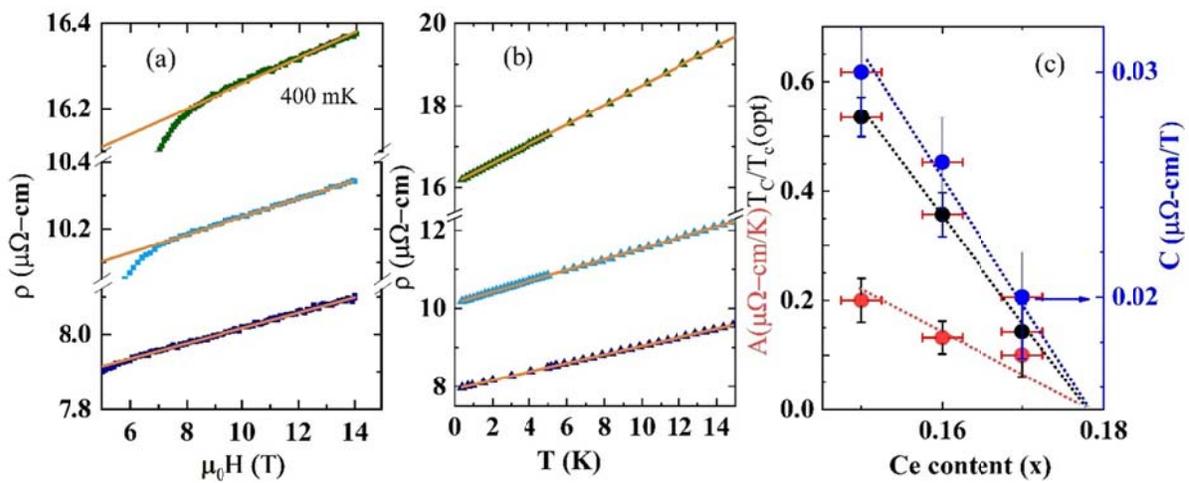

**Figure-2**

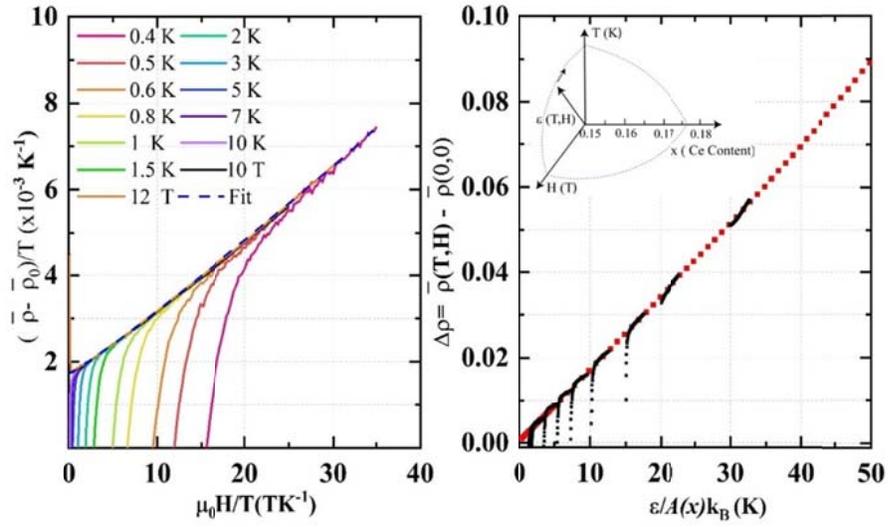

**Figure-3**

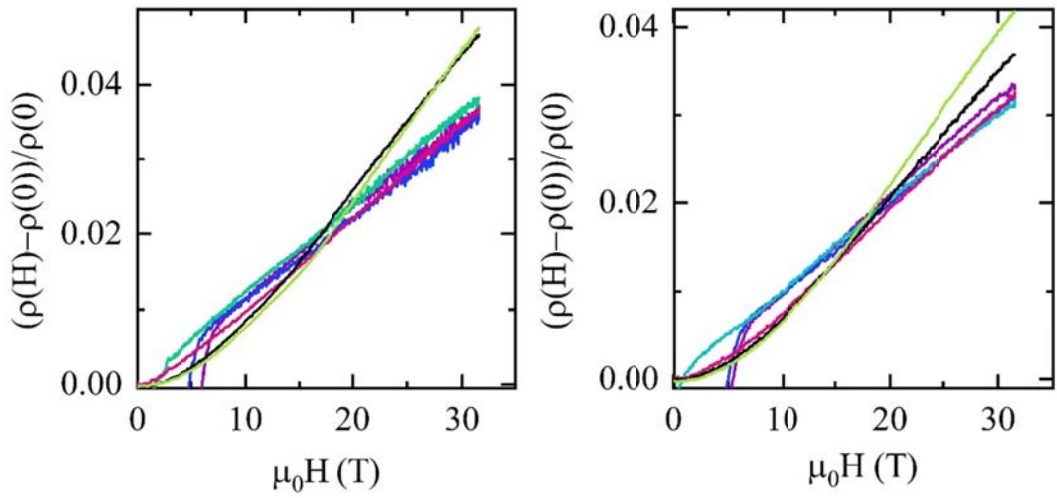

**Figure 4**

<br />

<br />

**Method:**

The measurements have been performed on $La_{2-x}Ce_xCuO_4$ (LCCO) films for $x$=0.13, 0.15, 0.16, 0.17, 0.18 compositions. High quality LCCO films (thickness about 150 - 200 nm) were grown using the pulsed laser deposition (PLD) technique on $SrTiO_3$ [100] substrates (5×5 mm$^2$) at a temperature of 750 °C utilizing a KrF excimer laser. The targets of LCCO have been prepared by the solid-state reaction method using 99.999% pure $La_2O_5$, $CeO_5$, and CuO powders. The Bruker X-ray diffraction (XRD) of the films shows the c-axis oriented epitaxial LCCO tetragonal phase. The thickness of the films has been determined by using cross sectional scanning electron microscopy (SEM). The resistivity measurements of the films have been carried out in the 400 mK to 200 K in DC magnetic fields up to ±14 T in a Quantum Design Physical Property Measurement System) with same pattern geometry for all the samples. The Hall component in the magnetoresistance is removed by adding positive sweep and negative sweep and dividing by 2. In some films the measurement is done up to 65 T. The high field 65 T measurement is done by standard four probe AC lock-in method at the National High Magnetic Field Laboratory (NHMFL) pulsed field facility, Los Alamos National Laboratory. The 31 T DC field measurement was done at NHMFL, Florida State University, Tallahassee.

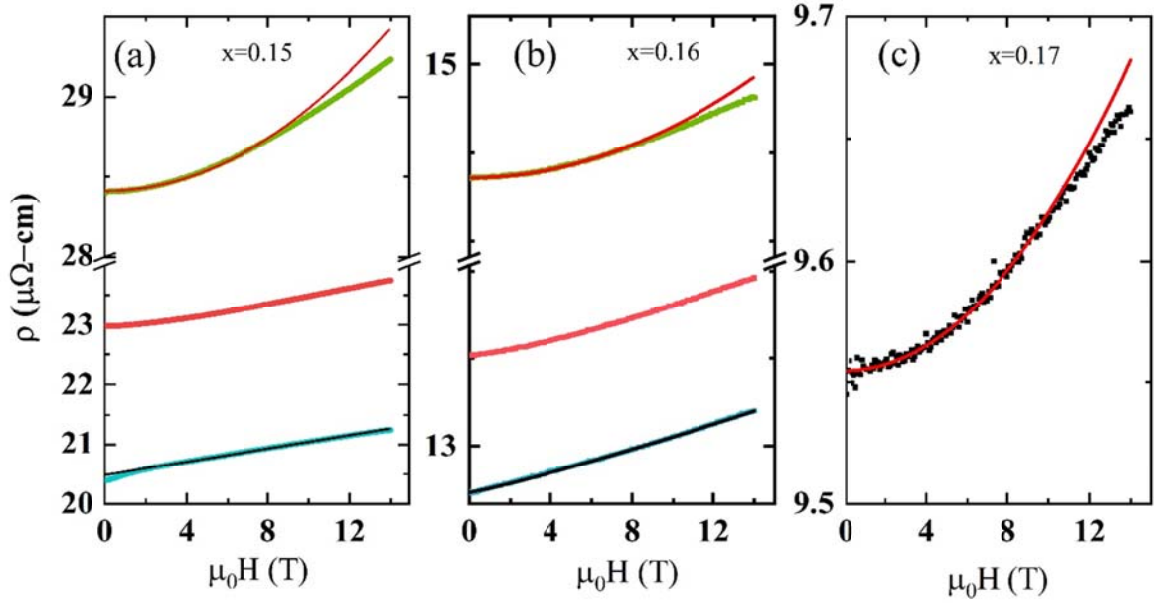

**Figure S1: Magnetoresistance vs doping:** ab-plane resistivity versus transverse magnetic field (H//c axis) as a function of temperature (color line) for x =0.15 (20 K, 30 K, 50 K) 0.16 (20 K,30 K,40 K) and 0.17 (15 K). The fit $\rho(H) = \rho(0) + C(x)(\mu_0 H)^n$ where n=1 for 0.15 (20 K and 1.13 for 0.16 (20 K)) (black solid line) and $\rho(H) = \rho(0) + K(x)(\mu_0 H)^2$ (red solid line) for x=0.15 (50 K), 0.16 (40 K) and 0.17 (15 K).

Figure S1 shows the magnetoresistance temperature crossover regime from linear-in-H to quadratic-in-H for dopings x=0.15, 0.16. The data shows that the crossover temperature decreases with increasing doping. Linear-in-H magnetoresistance exists to very low field for doping x=0.15 (20 K) and x=0.16 (20 K) and x=0.17 (~5 K see main text figure 1). The magnetoresistance in x=0.16 at 20 K is almost linear down to zero field which is clear from the fitting $\rho(H) = \rho(0) + C(x)(\mu_0 H)^n$ where n=1.13. That n is not exactly 1 could result from a small quadratic-in-H contribution. The magnetoresistance at higher temperature goes like $H^2$ as shown in figure S1 with fitting and in figure S2 measured up to 80 K. Thus, at higher temperature the transverse magnetoresistance is similar to that found in conventional metals. However, the origin of the higher T MR is yet to be determined.

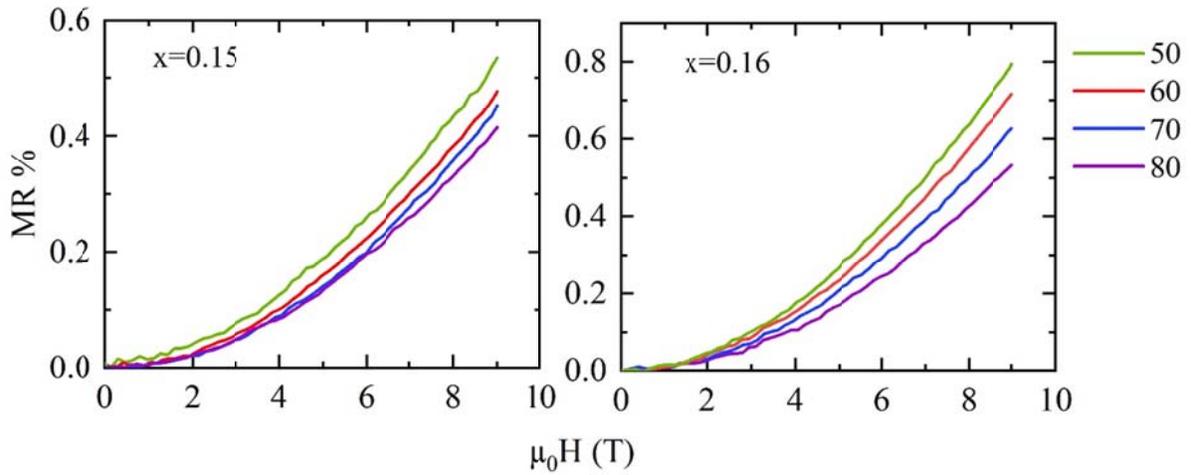

**Figure S2: Magnetoresistance vs doping above crossover:** (a), (b) transverse *ab* plane magnetoresistance MR% = $((\rho(H) - \rho(0))\times 100/\rho(0)$ for sample x=0.15 and 0.16 measured up to dc field of 9T for temperature 50 K to 80 K.

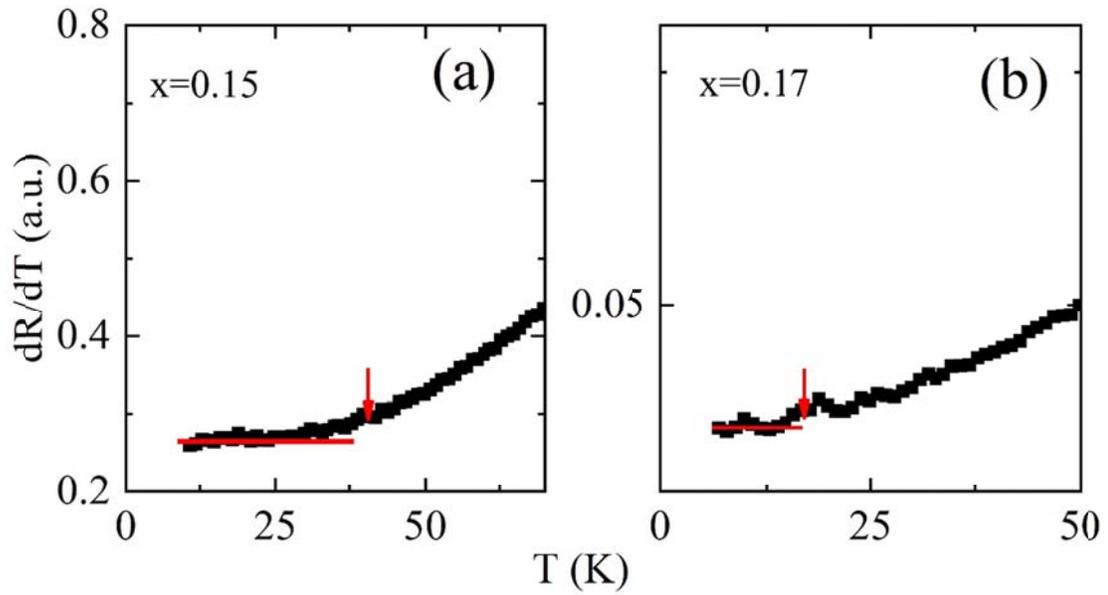

**Figure S3: Derivative of normal state resistance:** dR/dT of the resistance vs temperature, red arrow indicates the change in the slope for x=0.15 and 0.17 at H=0.

The change in the slope indicates the temperature where linear-in -T resistivity crosses over to quadratic-in-T resistivity. Here we have shown the cross over temperature ($T_\rho$) for two doping x=0.15 and 0.17. The evolution of the cross over temperature as function of doping is shown in figure S8.

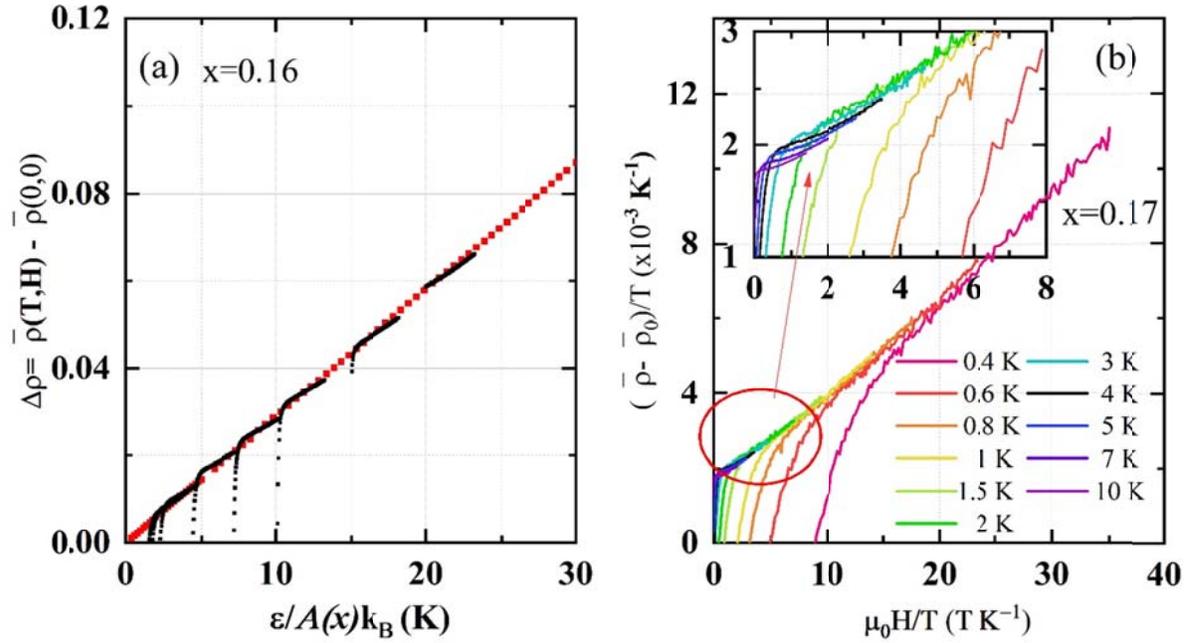

**Figure S4: Scaling between field and temperature:** (a) for x=0.16 $\bar{\rho}(T,H) - \bar{\rho}(0,0)$ vs $T + \frac{C(x)\mu_B}{A(x)k_B}\mu_0 H \equiv \varepsilon(T,H)/A(x)k_B$. $\bar{\rho}(0,0)$ is taken from extrapolating the zero field resistivity data down to T=0. Resistivity data (red) is the data with 7 T field after subtracting the $\bar{\rho}(0,7T)$. (b) for x=0.17 $(\bar{\rho} - \bar{\rho}(0))/T$ (where $\bar{\rho}$ $(= \rho(T)/\rho(200K))$, ($\bar{\rho}(0) = \frac{\rho(0, 0.4)}{\rho(0,200)}$ taken from figure 2a) vs $\mu_0 H/T$ is been deduced with varying temperature at fixed field as well as varying field at fixed temperature) (color solid lines) for doping x=0.17. Scaling plot of the transverse magneto resistivity is fitted with $\Delta\rho = \alpha + \beta(\mu_0 H/T)^\gamma$ ($\gamma$=1.09) (blue dashed line).

This anomalous low temperature scattering behavior cannot be explained by any orbital magnetotransport mechanism which will strongly depend on temperature and be quadratic in field (decreases with increasing temperature). We attribute this anomalous scattering to excitations associated with the QCP or excitations of unknown origin. The high temperature MR is quadratic with field. The scaling works only at low temperatures where magnetoresistance is linear-in-H. In the inset it is clearly shown that for x=0.17 samples the scaling is not working above 4 K where the magnetoresistance is deviating from linear-in-H. In contrast the scaling works to a much higher temperature for x=0.15 as linear-in-H exists to a higher temperature.

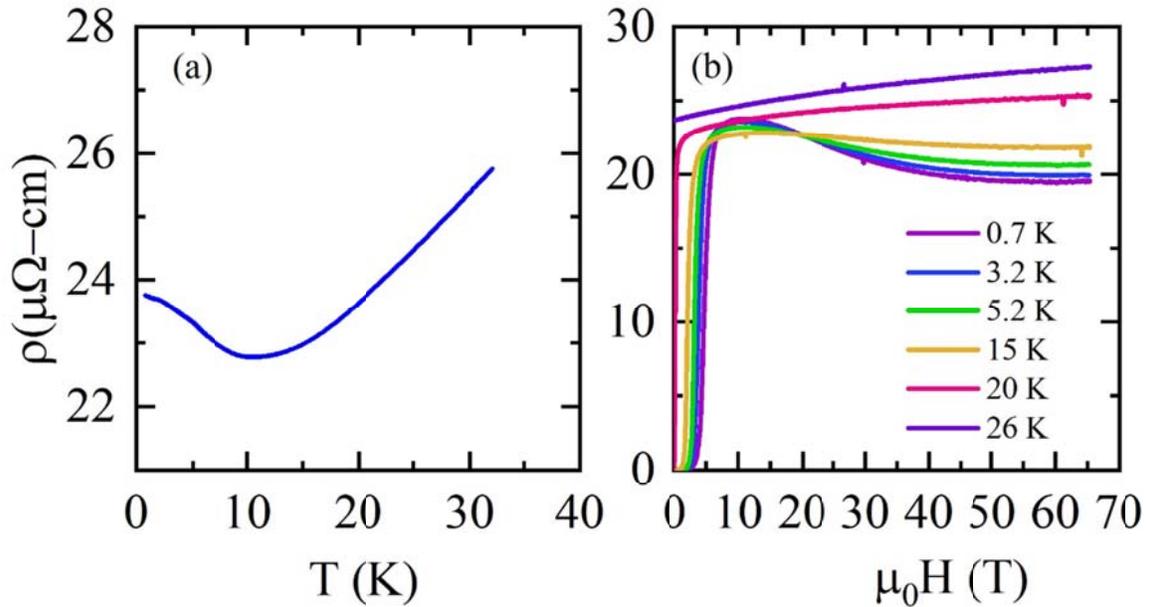

**Figure S5: Resistivity vs Temperature and field for x=0.13:** a) The normal-state ab-plane resistivity versus temperature in a magnetic field of $10\ T\ >\ H_{c2}$ applied parallel to the c axis. b) ab-plane resistivity vs magnetic field measured up to 65 T (pulsed field) down to 700 mK.

Fig-S5 shows the resistivity and MR for a doping just below the FSR (x<0.14). In electron doped cuprates the AFM order (long range or short range) vanishes at a critical doping $x_c$, where the low temperature normal state resistivity upturn also ends (x=0.14 in LCCO). In this study we find that the resistivity minimum is associated with negative transverse magnetoresistance as shown in figure S5. Once doping increases above $x_c$, the field driven normal state magneto resistivity is positive and linear-in-H (see main text).

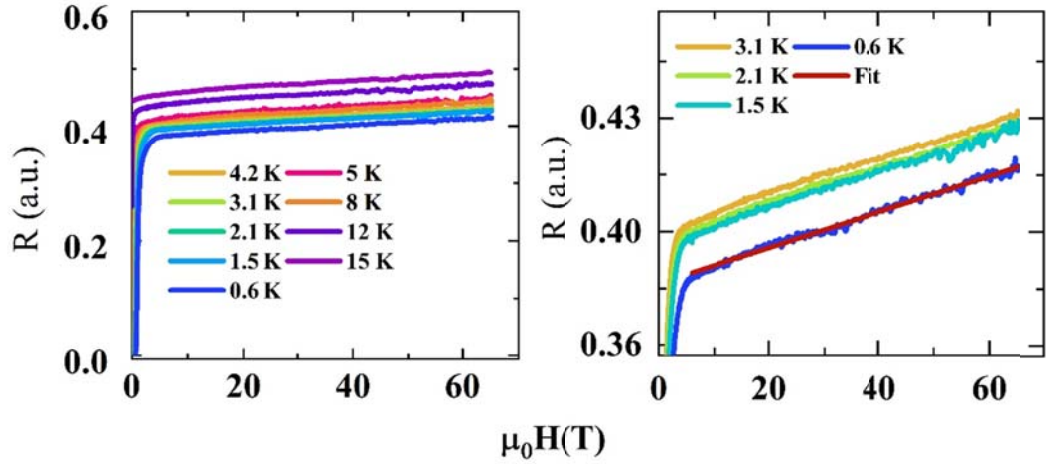

**Figure S6: High field magnetoresistance of x=0.15:** resistance vs magnetic field up to 65T for x=0.15 as a function of temperature (color solid line) with a fit (red solid line). $R(H) = R(0) + \mu_0 H$ at the lowest temperature 600 mK.

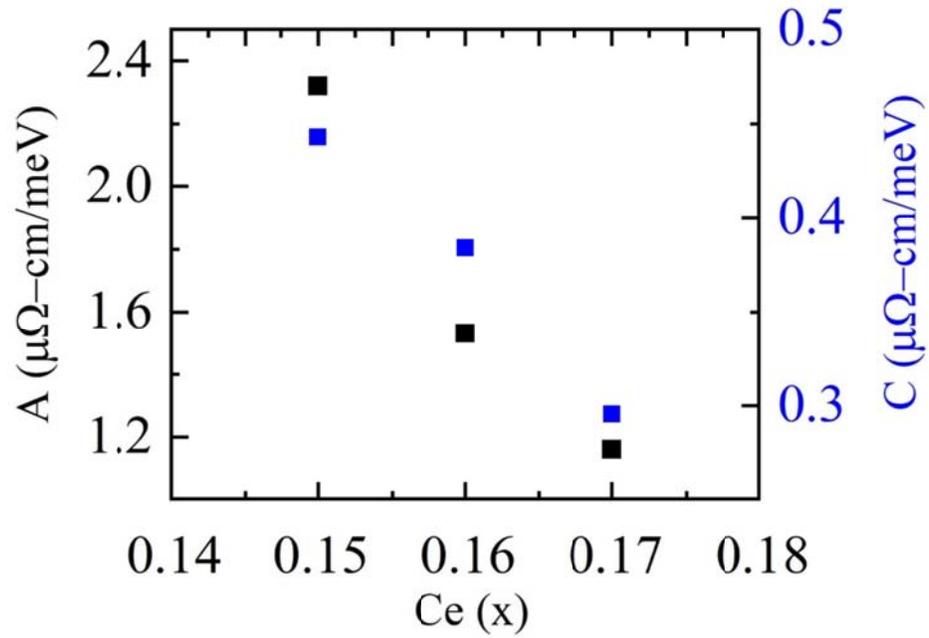

**Figure S7: Scattering rate vs doping:** Doping evolution of the scattering coefficients (temperature-slope (black) and field-slope (blue)) for x=0.15, 0.16 and 0.17. The values of the slopes are in energy units.

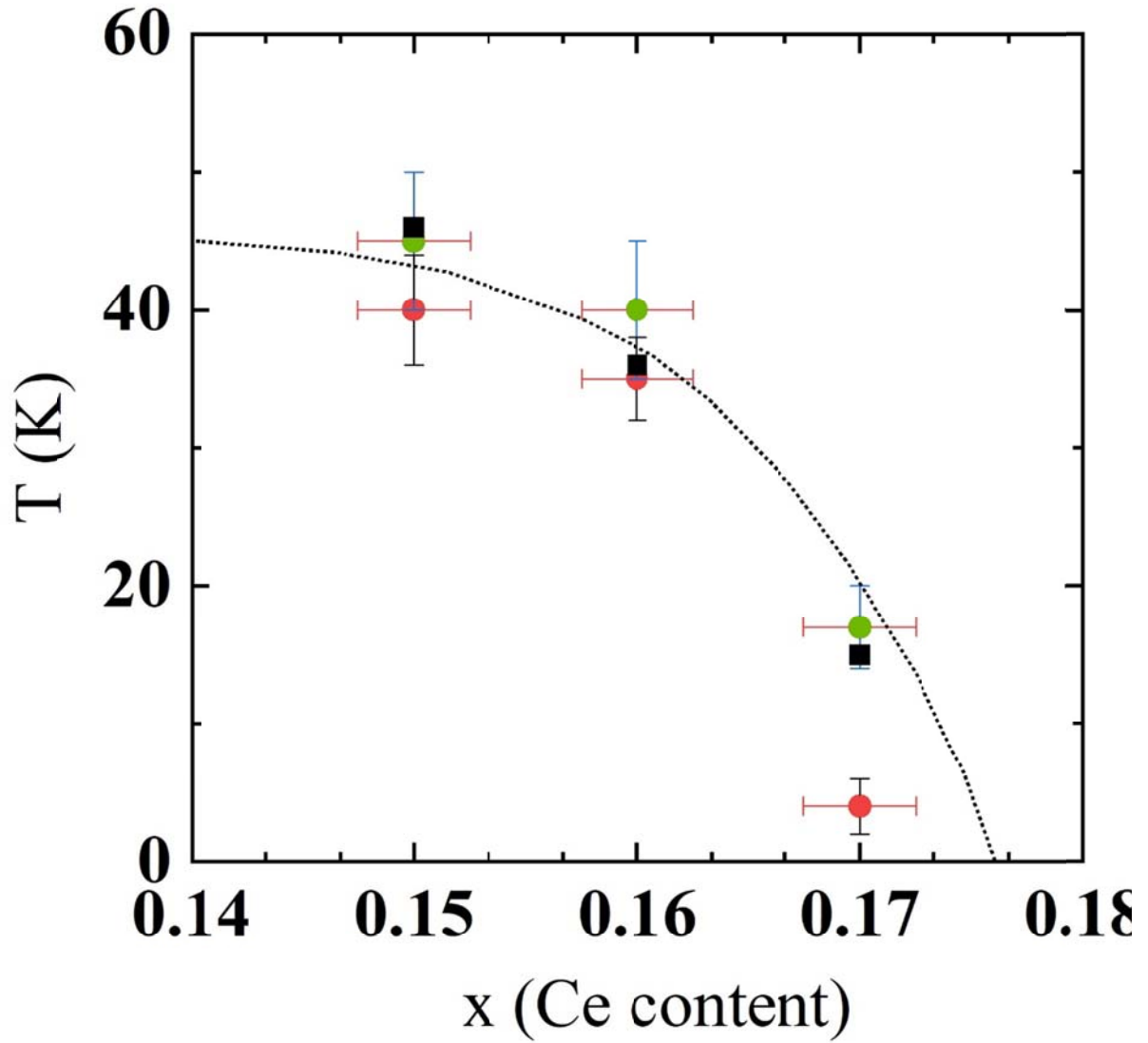

**Figure S8: Temperature vs doping phase diagram:** Temperature-doping phase diagram for x=0.15, 0.16 and 0.17 shows the linear-$H$ magnetoresistance (red)- temperature at which magnetoresistance goes from linear to quadratic (taken from Figure-1 and Fig-S1).), (green)- temperature at which *ab*-plane resistivity goes from linear to quadratic taken from Fig-S3 and $3T_c$ (black). Dotted line is the approximate temperature boundary below which $\rho \propto T, \rho \propto \mu_0 H$.

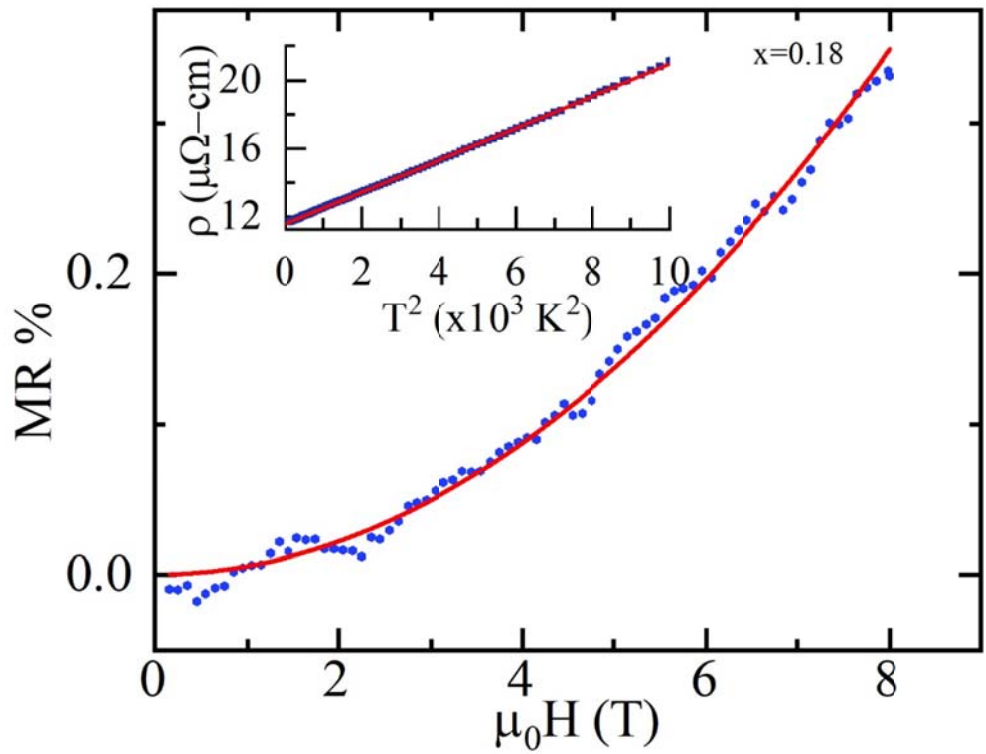

**Figure S9: Magnetoresistance for x=0.18:** Magnetoresistance at 5 K (H//ab plane) fitted by

MR% $= (\frac{\rho(H)-\rho(0)}{\rho(0)} 100) \propto (\mu_0 H)^2$ $(red)$; Inset: resistivity ρ(T) vs $T^2$ (blue).